# Biomechanical Influence of Cavity Fillers on Peri-wound Soft Tissue Strains in Sacral Pressure Injuries

Coraline Staub-Milants[1,2*], Grégory Chagnon[1], Noémie Briot[1], Laurent Pasquinet[2], Yohan Payan[1]
[1] *Université Grenoble Alpes, CNRS, UMR 5525, VetAgro Sup, Grenoble INP, TIMC, Grenoble, France*
[2] *Urgo Research, Innovation and Development (URID), Chenôve, France*
* *Corresponding author: staubc@univ-grenoble-alpes.fr*

**ABSTRACT**

**Aims**: Deep cavity pressure injuries (PIs) are commonly managed with wound fillers, which biomechanical role remains poorly understood. This study aimed to assess how filler mechanical properties influence peri-wound soft tissue strains.

**Methods**: This study combines hydrated mechanical testing of commercial cavity fillers with finite element modeling of a sacral stage 3 PIs to assess how filler-related parameters affect peri-wound soft tissue shear strains.

**Results**: Sensitivity analysis shows that filling ratio is the dominant factor governing the volume of mechanically preserved tissue, outweighing filler stiffness, friction, and compressibility. High filling ratios markedly reduce strain concentrations, while more compliant, adipose-like fillers further improve tissue protection.

**Conclusion**: These findings highlight the importance of dead-space elimination and tissue-matched filler design.

**Declaration of interest :** This work was supported by the French National Centre for Scientific Research, France, and Urgo Research Innovation & Development (URID), France. CSM and LP are employed by URID. URID provided financial support for the study and reviewed the manuscript prior to submission. The authors declare that the interpretation of the data and the conclusions presented in this manuscript are their own. The remaining authors have no conflict of interest to declare

**KEY POINTS**

- The wound filling ratio was identified as the dominant parameter influencing peri-wound soft tissue protection, accounting for more than 98% of the variance in mechanically preserved tissue volume.
- More compliant fillers with adipose-like mechanical behaviour further improved tissue protection by reducing strain transfer to surrounding tissues.
- The findings support current recommendations advocating dead-space elimination in cavity wound management and highlight the potential value of tissue-matched, highly compliant wound fillers.

## 1. Introduction

In the management of deep cavity wounds, such as stage 3 and 4 pressure injuries (PIs), wound fillers are commonly used to occupy the wound volume. Their functions include eliminating dead space, maintaining contact with the wound bed, and managing exudate while preserving a moist healing environment.[1] A broad array of fillers exists (hydrogels, hydrocolloids, foams, alginates, hydrofibers),[2] which makes their selection challenging in clinical practice.[3,4]

To guide clinical decision-making, numerous in vitro studies have characterized the absorption capacity, retention properties, and swelling kinetics of these materials.[5–8] In contrast, their mechanical behavior has received comparatively little attention, despite strong evidence linking pressure ulcer development to shear strains within soft tissues.[9–11] Moreover, mechanical properties are predominantly assessed on dry dressings,[7,12] although the few studies performed under hydrated conditions show marked changes in stiffness.[6,13] To the authors' knowledge, only one modelling study has considered the effect of a wound filler in a cavity wound, namely Fougeron et al.[14], who
showed that the presence of a filler significantly affects tissue strain fields. However, the filler effect was approximated as a uniform cavity pressure and not represented explicitly. In parallel, several studies have examined the mechanical influence of dressings applied at the skin surface, highlighting the critical role of friction conditions[15–17] and material stiffness in modulating internal tissue stresses.[18] Nevertheless, these approaches do not explicitly represent a material filling the wound cavity.

In this context, the present study aims to quantify how cavity filler-related parameters influence peri-wound soft tissue shear strains in a finite element model of a sacral cavity wound in which the cavity filler is explicitly represented as a hyperelastic material. More specifically, the analysis focuses on the effects of filling ratio, filler mechanical response, compressibility, and friction at the filler-tissue interface. To support a plausible range of filler mechanical behaviors, hydrated tensile and compression tests were performed on several commercial cavity fillers.

## 2. Materials and Methods

### *2.1. Model Description*

The model used in this study was adapted from that of Fougeron et al..[14] It represents an open cavity wound beneath the sacrum, including the bony prominence, and comprises the skin, adipose tissue, wound filler, secondary dressing, and mattress supporting a supine patient (see Figure 1). The 10 mm wound depth is assumed to represent a stage III pressure ulcer, commonly treated with cavity filler.

According to Fletcher et al.,[19] clinical wound packing varies, with material densely packed or layered. To reflect this practice while limiting simulated contact interfaces, the filler was modeled as a solid occupying 70 % to 100 % of the cavity volume and contacting both the wound bed and secondary dressing.

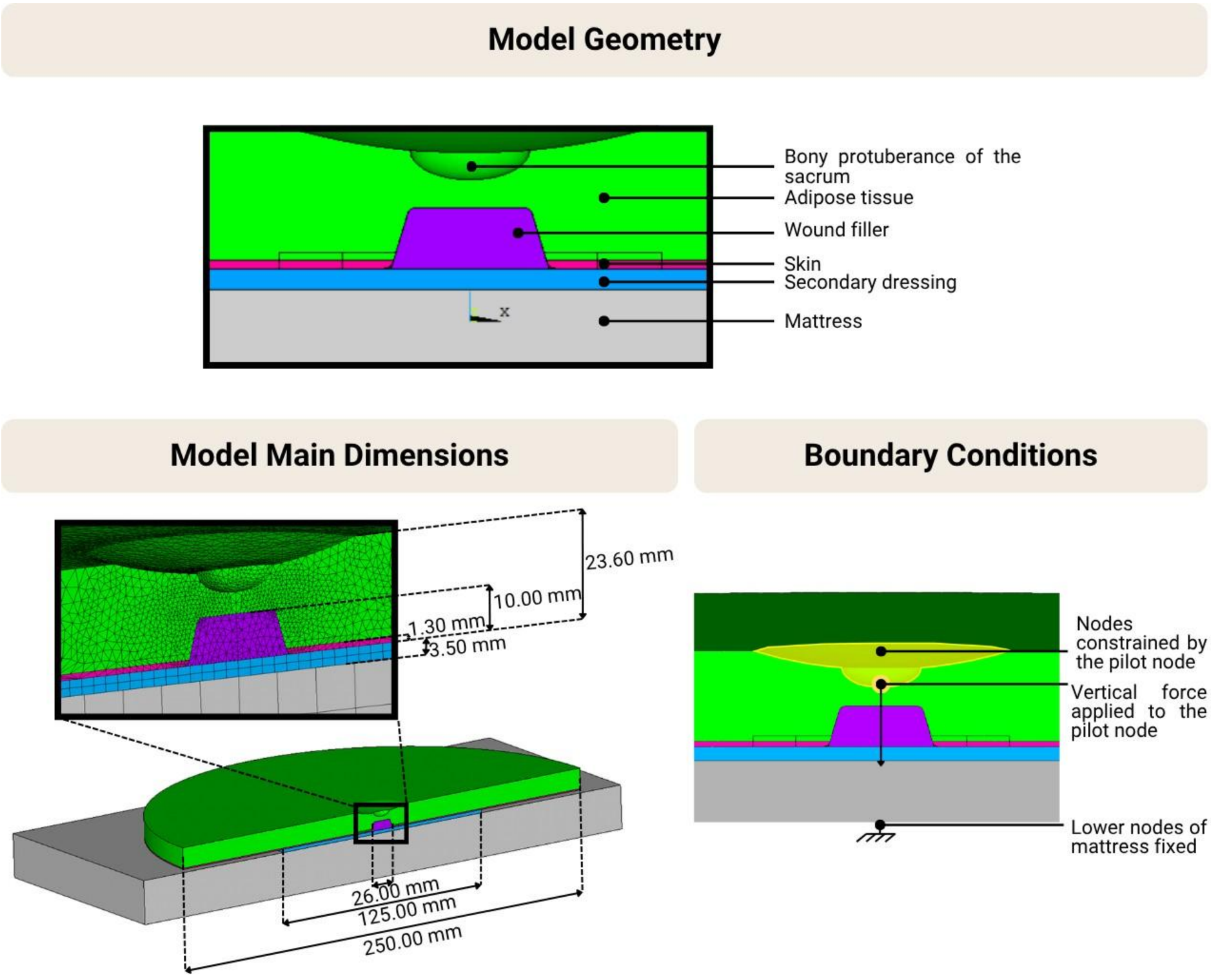

**Figure 1.** Finite Element Model (adapted from that of Fougeron et al.[14])

## 2.2. *Mechanical Characterization of Materials*

### 2.2.1. *Tensile and Compression Tests*

To define the mechanical properties of the cavity filler in the numerical model, uniaxial tensile and unconfined compression tests were performed on several commercial wound dressings. The selected products represent the main fiber technologies used for cavity fillers, namely polyacrylate, alginate, polyvinyl alcohol (PVA) and carboxymethylcellulose (CMC). The tested dressings were UrgoClean Rope (Urgo), Algostéril (Brothier Laboratoire), Exufiber (Mölnlycke), and Aquacel (Convatec). These tests were intended to inform the finite element model and identify representative hydrated filler behaviors, rather than to provide an exhaustive comparison of commercial products.

All tests were conducted using an MTS (Materials Test Systems) testing machine (model C42 503, ±25 N load cell, 0.01 % accuracy). To represent the worstcase absorption scenario, all samples were tested in a fully saturated state. According to the NF EN 13726 (Part 3.2 – September 2002) standard for wound dressing test methods, samples were immersed for 30 min in a $NaCl–CaCl_2$ solution with an ionic composition representative of wound exudate. The tests were then conducted with the samples immersed in the same solution bath maintained at 37 ± 0.5 °C, corresponding to the upper temperature range reported in the wound bed.[6,7,20]

Uniaxial tensile tests were performed on samples measuring (60.0 ± 0.5) mm × (7.0 ± 0.5) mm. Unconfined compression tests were performed by stacking several saturated samples to obtain a thickness of at least 10 mm. The stacked samples had a minimum surface area of 40 mm × 40 mm and were compressed using a circular platen (diameter 30 mm) until the sensor's upper limit.

All tests were performed at a displacement rate corresponding to a nominal strain rate of 1%/*s*. For each dressing type, a representative curve was selected for constitutive fitting.

#### 2.2.2. Volumetric Compressibility Tests

To capture realistic volumetric behavior, the pressure-volume response of an UrgoClean gauze assembly saturated with $NaCl$–$CaCl_2$ solution was measured using the device of Briot et al..[21] A cylindrical chamber containing stacked saturated gauze layers was piston-compressed. The response yielded the bulk modulus $K$, used to define the reference compressibility parameter $d$ of the homogenized finite element filler.

### 2.3. Numerical Implementation

The numerical simulations were performed using a finite element model implemented in ANSYS APDL (ANSYS 2023 R2, ANSYS Inc., Canonsburg, PA, USA) with an implicit quasi-static solver.

#### 2.3.1. Discretization

Linear hexahedral elements were used for the mattress and secondary dressing, whereas the soft tissues and wound filler were modeled with linear tetrahedral elements under a mixed pressure–displacement formulation. Mesh convergence was achieved at 42,062 elements in the perilesional region of interest, with relative differences below 1.5% for nodal displacement and 0.3% for the tissue preservation metric compared with the finest mesh (863,300 elements).

#### 2.3.2. Boundary Conditions and Contact Interactions

A vertical load was applied to a pilot node located in the sacral region (Figure 1), corresponding to 47 % of the body weight of a 94 kg subject to represent a supine subject, following Fougeron et al..[22] To reduce computational cost, only half of the geometry was modeled using a symmetry plane. The bottom nodes of the mattress were fully constrained.

Contact interactions were modeled with surface-to-surface elements using the augmented Lagrange formulation. Adipose tissue-skin and skin-secondary dressing interfaces were perfectly bonded. A friction coefficient of 0.62 was used between the secondary dressing and mattress, following Fougeron et al..[14] Filler-soft tissue friction was varied parametrically, while filler-secondary dressing friction was set to 0.9. In the absence of dedicated tribological data for this interface, this value was chosen as a conservative worst-case estimate based on available measurements for dry internal layers of multilayer prophylactic dressings (COF: 0.78 to 0.98[23]) and saturated dressing-skin contacts (COF: 0.15 to 1.5[17,24]). Its influence was also expected to be limited under predominantly vertical loading.

### 2.3.3. *Constitutive Equations*

Material models were implemented following Fougeron et al..[22] Skin and adipose tissues were modeled as hyperelastic, whereas the mattress and dressing layer were assumed linear elastic. For the wound filler, several constitutive laws and compressibility levels were tested in the sensitivity analysis (Section 2.4). The filler constitutive behavior was defined by separating its deviatoric and volumetric responses. The deviatoric response was described using a second-order Ogden model fitted in MATLAB R2024a (lsqcurvefit, MathWorks, Natick, MA, USA) to the experimental compression curves over the deformation range considered in this study, with stretch ratio $\lambda \in [0.75,1]$, due to the asymmetry between the compressive and tensile responses. The volumetric response was introduced independently through the compressibility parameter $d = \frac{2}{K}[Pa^{-1}]$, where $K$ is the bulk modulus, so that larger $d$ values correspond to higher compressibility. In the following, "material properties" refers to the Ogden parameters governing the deviatoric response, whereas "compressibility" refers only to the volumetric parameter.

The material parameters used in the simulations are summarized in Table 1.

**Table 1.** Material parameters used in the simulations

| **Material** | **Constitutive law** | **Parameters** | **Source** |
|---|---|---|---|
| Skin | Isihara[25] | $C_{10} = 2.7 \times 10^{-1}$ $MPa$ $C_{20} = 1.9$ $MPa$ $d = 7.5 \times 10^{-4}$ $MPa^{-1}$ ($v = 0.4999$) | Ni Annaidh et al.[26] |
| Adipose tissue | Yeoh[27] | $C_{10} = 1.3 \times 10^{-4}$ $MPa$ $C_{20} = 0.0$ $MPa$ $C_{30} = 12.2 \times 10^{-3}$ $MPa$ $d = 1.54$ $MPa^{-1}$ ($v = 0.4999$) | Sommer et al.[28] |
| Secondary Dressing | Linear elastic orthotropic | $E_x = 4.4$ *MPa* $E_y = 1.8$ *MPa* $E_z = 2.6 \times 10^{-2}$ *MPa* $d_{eq} = 1.48$ $MPa^{-1}$ ($v = 0.2560$) | Fougeron et al.[22] |
| Mattress | Linear elastic isotropic | $E = 0.23$ *MPa* $d = 10.4$ $MPa^{-1}$ ($v = 0.3000$) | Fougeron et al.,[14] Lee et al.[29] |
| Filling Material | Yeoh[27]/Ogden | Varies according to the sensitivity analysis | Experimental tests (Part 2.2) |

### 2.4. *Sensitivity Analysis*

A sensitivity analysis based on the method of Macron et al.[30] was performed to identify the wound filler parameters that most influence the volume of mechanically preserved soft tissue. Maximal Green-Lagrange shear strain ($\epsilon_{shear}$, Eq. 1) is a recognized mechanical biomarker of soft tissue viability and pressure ulcer risk.[9] Sensitivity was therefore quantified from the soft tissue volume within a perilesional region of interest (ROI), defined as a radius of 1.2 times the wound radius, subjected to maximal shear strain below 0.3. This threshold was previously used in pressure ulcer biomechanics sensitivity analyses[14] and is supported by rat experiments suggesting that strains above 0.3 are associated with tissue damage.[9] The ROI was restricted to the peri-wound region, where filler effects were expected to be the highest.

$$\epsilon_{shear} = \frac{1}{2}\max(|\epsilon_1 - \epsilon_2|, |\epsilon_1 - \epsilon_3|, |\epsilon_2 - \epsilon_3|) \text{ with } \epsilon_i = \frac{1}{2}(\lambda_i^2 - 1), \mathrm{i} \in [1,2,3] \quad (1)$$

with $\epsilon_i$ the Green-Lagrange principal strains.

To address clinically relevant questions, the following parameters were investigated in the sensitivity analysis:

- **Wound filling ratio:** discussions with clinicians indicated an ideal wound filling ratio of approximately 80 %, ensuring contact between the filler, wound bed, and secondary dressing to promote exudate drainage and adipose tissue regeneration. The ratio was varied from 70 % to 100 % to reflect clinical variability.
- **Wound filler material properties:** Material 1 was defined as a hypothetical very compliant filler with properties identical to those of adipose tissue and used as a reference to assess the benefits of a more tissue-like response. Based on the experimental tests (Part 2.2), the most rigid dressing (Exufiber) was selected as Material 3 and an intermediate material is selected as Material 2.

Figure 2 summarizes the experimental results and the material choices (Materials 1–3).

- **Compressibility coefficient:** as the compressibility parameter *d* of saturated UrgoClean gauze is estimated at 0.0007 $MPa^{-1}$, it was varied from 0.0007 to 0.07 $MPa^{-1}$ to account for effective compressibility from material properties and exudate saturation.
- **Friction coefficient between the cavity filler and soft tissues:** friction at this interface is expected to depend on body fluids,[17] wound moisture,[17,24] and filler material properties.[24] To the authors' knowledge, no direct tribological data are available for the cavity filler-soft tissue interface. The investigated range (0.1 to 1.1) was therefore defined as an exploratory interval based on dressing-skin reported values (0.15 to 1.5[17,24]), excluding the highest values due to the different hydrated internal interface considered here.

Parameter ranges are summarized in Table 2, leading to a total of 81 simulations covering all combinations. The volume of mechanically preserved soft tissue in the peri-wound region was then modeled using a full quadratic response surface based on normalized parameters, with coefficients fitted on the converged simulations.

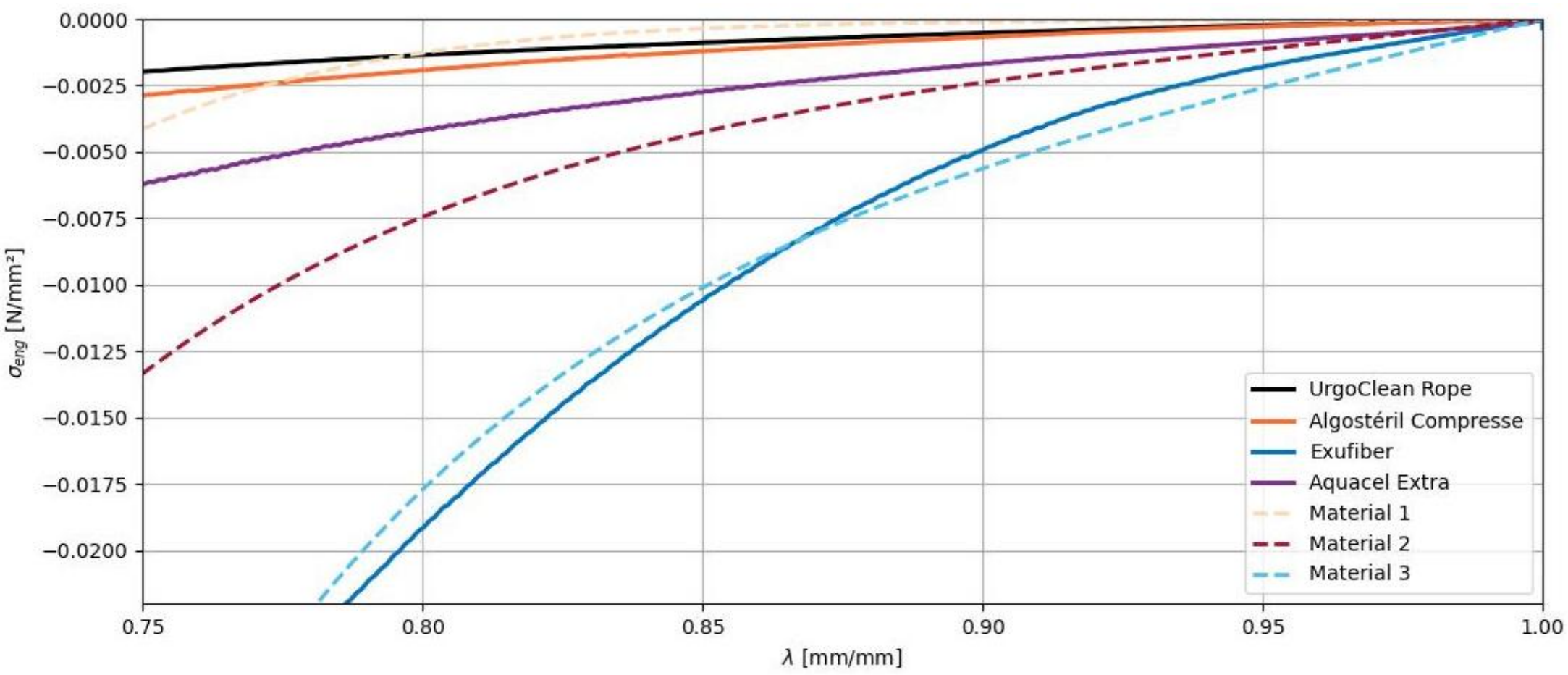

**Figure 2.** Engineering stress, σeng, versus stretch ratio, λ, for compression tests using a two-term Ogden model[39] for Materials 2 and 3: Material 1 (adipose tissue, refer to Table 1); Material 2 (μ1=0.01 MPa, α1=0.01, μ2=0.001 MPa, α2=16); Material 3 (μ1=0.1135 MPa, α1=0.0989, μ2=0.0013 MPa, α2=18.6946). Solid lines: experiments; dashed lines: model fits. Material 1—hypothetical very compliant filler with adipose tissue stiffness properties; Material 2—intermediate filler; Material 3—most rigid filler among the fillers tested. UrgoClean Rope (Urgo Medical, France); Algostéril Compresse (Brothier Laboratoire, France); Exufiber (Mölnlycke Health Care, Sweden); Aquacel Extra (Convatec, UK)

**Table 2.** Minimum, intermediate and maximum parameter values in sensitivity analysis

| **Parameters** | **Minimal Level** | **Intermediate Level** | **Maximal Level** |
|---|---|---|---|
| Wound filling ratio | 70 % | 85 % | 100 % |
| Material properties | Material 1 | Material 2 | Material 3 |
| Wound filler compressibility | 0.0007 $MPa^{-1}$ | 0.007 $MPa^{-1}$ | 0.07 $MPa^{-1}$ |
| Friction coefficient Soft tissues-Wound filler | 0.1 | 0.6 | 1.1 |

## 3. Results

The model demonstrates an excellent fit and predictive capability ($R^2 = 0.9985$, leave-one-out $Q^2 = 0.9976$). The regression matrix condition number of 6.35 indicates good numerical stability. Only statistically significant parameters ($p < 0.05$) are reported.

The wound filling ratio is by far the main contributor to the variance of mechanically preserved soft tissue volume (98.2 %), followed by material properties (1.71 %) and friction coefficient (0.133 %) while compressibility has no significant influence within the near-incompressible range investigated. The signs of the regression coefficients suggest that an optimal filling configuration requires a high filling ratio, a compliant wound filler, and a high friction coefficient. Furthermore, the sensitivity analysis suggests that the influence of the friction coefficient on soft tissue viability is amplified at lower filling ratios and with softer cavity filler material. This influence remains weak, with sensitivities of 0.0647 % and 0.00910 % for the corresponding interaction terms.

When restricting the sensitivity analysis to simulations corresponding to the optimal filling ratio (100 %), the metamodel continues to accurately approximate the quadratic response surface, as indicated by $R^2 = 0.9939$ and $Q^2 = 0.9848$. The relative contributions are then dominated by material properties (92.4 %), followed by the soft tissue–gauze friction coefficient (7.78 %), while the contribution of the compressibility remains negligible. A key observation is that lower friction coefficients are beneficial for the fully filled configuration but detrimental in the other filling ratio scenarios.

Figure 3 illustrates the importance of the wound filling ratio and the stiffness of the cavity filler on the volume of mechanically preserved soft tissue. Visually, it is clear that reducing the stiffness of the material at a constant filling ratio is important, despite the relatively minor impact of material properties compared to the filling ratio. The ratio of mechanically preserved soft tissue volume to total soft tissue volume in the ROI was 0.71±0.063 for a 70 % filling ratio, 0.89±0.031for a 85% filling ratio, and 0.93±0.018 for a 100 % filling ratio. For comparison, the simulation without any filling (black curve on Figure 3) yields a mechanically preserved soft tissue ratio of 0.33. These findings indicate that increasing the filling ratio markedly improves the proportion of mechanically favorable tissue, with gains exceeding 50 % relative to no filling.

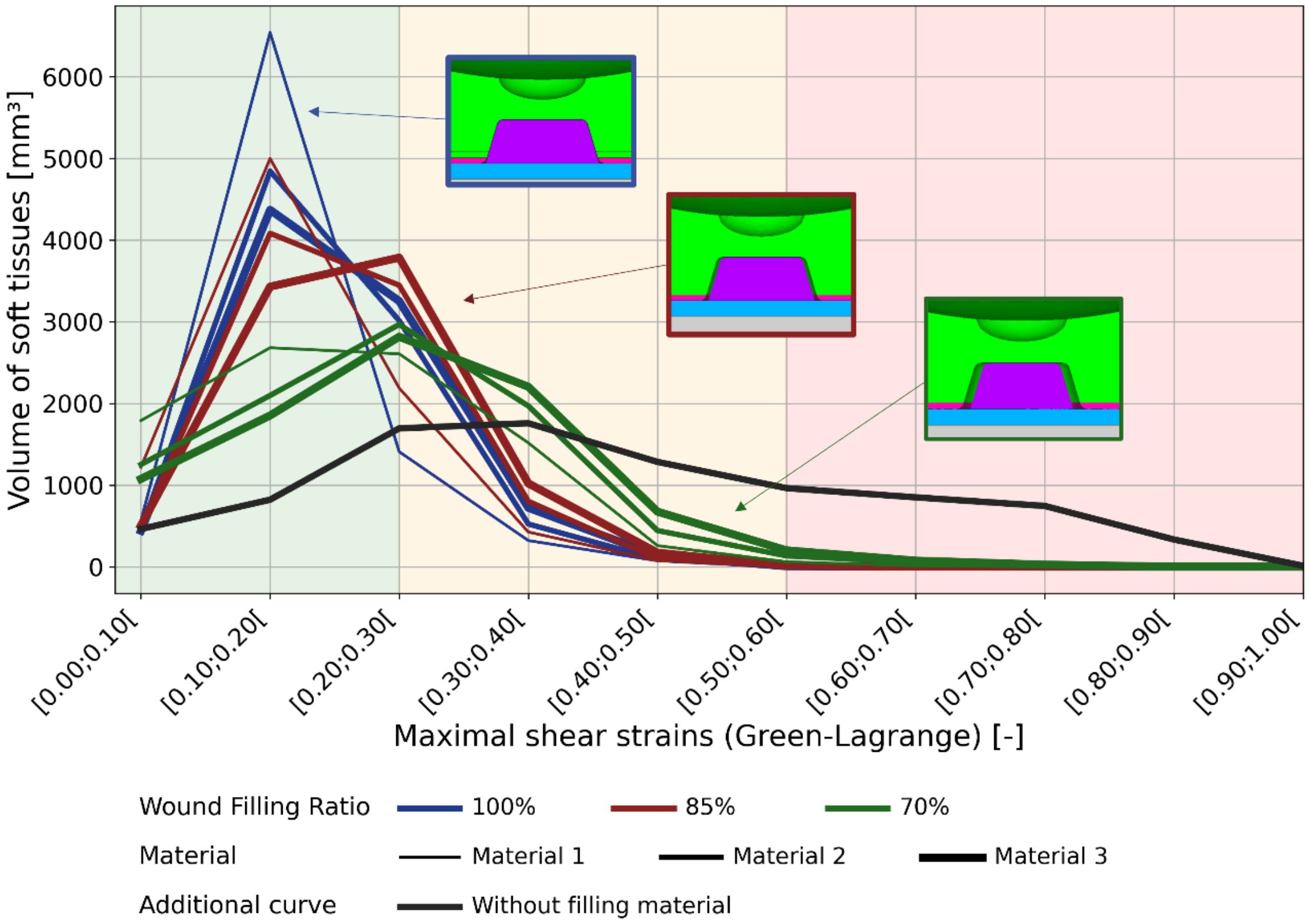

**Figure 3.** Effect of wound filling ratio (a: 100%; b: 85%; c: 70%) and material properties on mechanically preserved soft tissue volume in the region of interest for volumetric compressibility parameter d=0.007 MPa and friction coefficient 0.6. Green, yellow and red backgrounds indicate safe, potentially injurious and injurious strain regions, respectively.[9] Material 1—hypothetical very compliant filler with adipose tissue stiffness properties; Material 2—intermediate filler; Material 3—most rigid filler among the fillers tested

Figure 4 illustrates the effect of wound filling ratio on the distribution of maximal shear strains. Wound filling prevents the surrounding soft tissues from collapsing onto themselves under the load of a supine patient positioned over a sacral wound. Consequently, increasing the filling ratio reduces tissue deformation and leads to lower shear strain levels. Contact pressure distributions at the soft tissue surface further show that adequate filling improves load redistribution across the wound and reduces local pressure peaks. However, the presence of potentially critical shear strain ($\epsilon_{shear} > 0.3$) in the completely filled configuration indicates that the filling ratio alone is insufficient to ensure tissue viability under the conditions investigated.

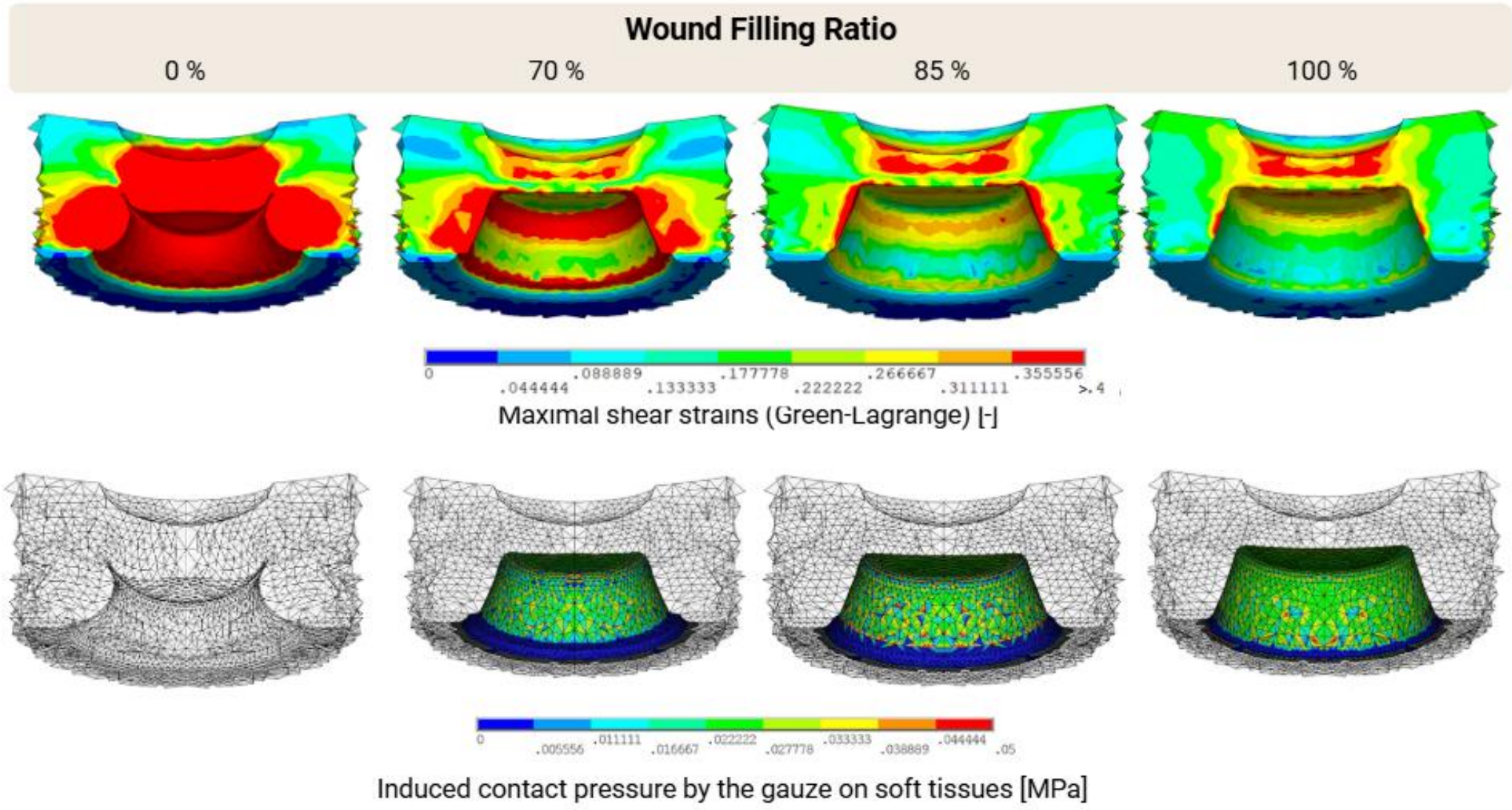

**Figure 4.** Impact of wound filling ratio on maximal shear strain and contact pressure distribution in the region of interest. The potentially damaging Green–Lagrange shear strains are represented in orange–red. The other parameters (coefficient of friction, material properties and wound filler compressibility) are set to their intermediate value

As already demonstrated in Figure 3, the mechanical properties of the wound filler also have a significant influence on the proportion of mechanically preserved soft tissue. Figure 5 shows that more compliant materials accommodate more deformation, reducing the mechanical load on surrounding soft tissues under vertical loading. Consequently, even at an 85 % filling ratio, a filler with adipose-like properties preserves 95 % of soft tissue mechanically.

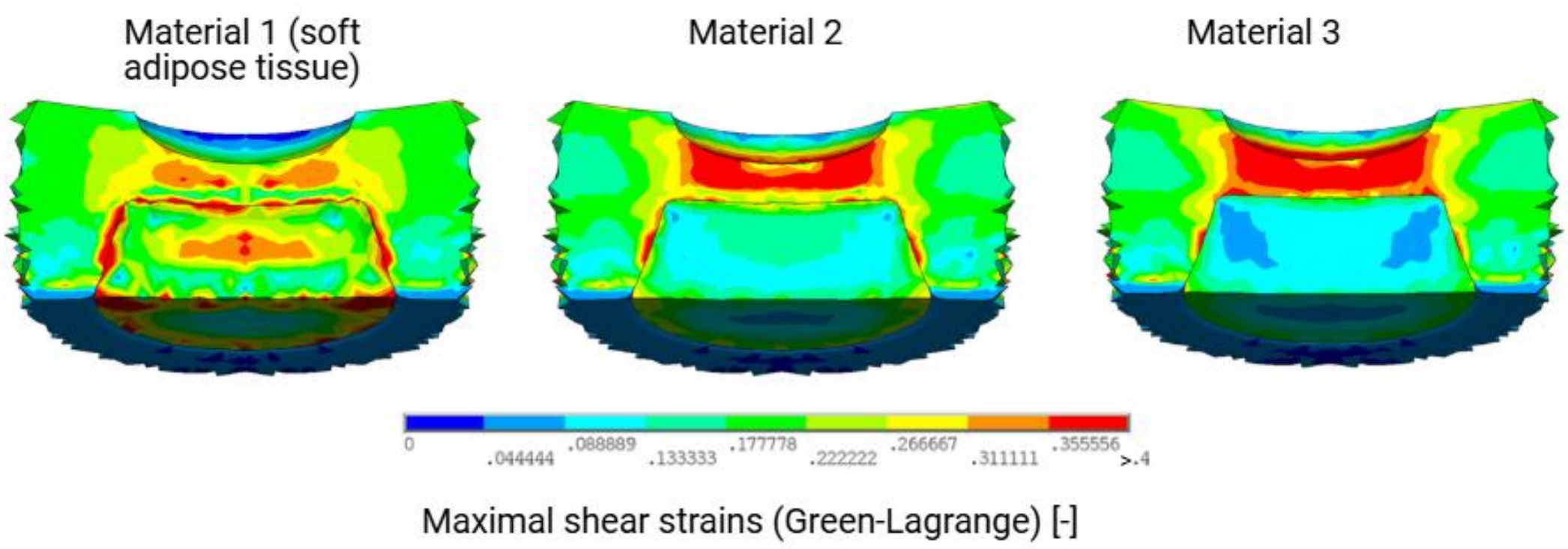

**Figure 5.** Impact of material properties on maximal shear strain distribution in the region of interest. The potentially damaging Green-Lagrange shear strains are represented in orange-red. The other parameters (wound filling ratio, coefficient of friction, wound filler compressibility) are set to their intermediary value. Material 1 – hypothetical very compliant filler with adipose tissue stiffness properties ; Material 2 – intermediate filler ; Material 3 – most rigid filler among the fillers tested

## 4. Discussion

Among the investigated parameters, wound filling ratio had the strongest influence on the peri-wound mechanical environment. Increasing filling ratio consistently improved tissue mechanical conditions, although filling alone was insufficient to preserve surrounding soft tissues mechanically. Filler stiffness also affected load transfer, with more favorable conditions obtained when the filler response was closer to that of adipose tissue. These results indicate that both adequate cavity filling and mechanical compatibility contribute to protection. The beneficial effect of a high filling ratio may be explained by improved geometric continuity, which promotes more homogeneous load transfer and reduces local stress and strain concentrations at the wound edge. Conversely, incomplete filling creates structural discontinuities that increase deformation gradients in peri-wound tissues. This interpretation agrees with EPUAP and WUWHS recommendations advocating dead-space elimination,[1,32] and with biomechanical studies showing that reducing stiffness mismatch between dressings and soft tissues improves load redistribution.[33]

The predicted mechanical levels were consistent with the literature. Interfacial pressures ranged from 0 to 0.04 MPa, which is consistent with previous reports.[14] Green-Lagrange shear strains reported in sacral finite element models range from about 0.2 in low-loading regions to 0.8 to 1.0 near bony prominences,[34] with similar magnitudes described in studies reporting principal strains.[35–37] The values obtained here (0.15 to 0.55) are broadly consistent with these previously reported values, although the upper end of the range observed here remains below the highest strains reported near bony prominences.

The influence of friction was more limited and depended on filling configuration. Compressibility shows no measurable effect on the output, despite a 100-fold variation across the studied interval, which is intended to capture hydrated dressing states.[38] However, the present results do not exclude a stronger effect for more compressible fillers.

Several limitations of the present model should be acknowledged. First, the filler was modeled as a homogeneous and isotropic material. This simplification enabled a systematic parametric analysis while avoiding complex contact interactions and numerical artefacts. It was considered appropriate because the study focused on the mechanical response of the surrounding tissues rather than on deformation within the filler microstructure. It may also be viewed as a conservative approximation with respect to stress transmission to adjacent soft tissues. Second, only the supine position was considered, as this posture corresponds to the condition most frequently maintained by patients at risk of developing such wounds. However, this assumption may lead to an underestimation of frictional effects. Given the potential influence of friction highlighted in this study, future work should investigate additional body positions to better capture the role of interface mechanics under more realistic loading scenarios. Finally, the model was limited to a single wound geometry and did not account for biological processes such as tissue remodeling or exudate production. In addition, no direct experimental validation of the cavity wound model was performed. The present framework should therefore be interpreted primarily as a comparative and mechanistic tool to assess the relative influence of filler-related parameters, rather than as a patient-specific predictive model.

## 5. Conclusion

The results identify wound filling ratio as the main driver of peri-wound mechanical conditions in the modeled configuration. However, favorable outcomes also require mechanical compatibility between the filler and surrounding tissue. These findings support the biomechanical importance of dead-space elimination and highlight the interest of highly deformable, tissue-matched wound fillers for deep wound management. Future work should incorporate time-dependent effects, evolving material properties, and fluid transport to better represent wound healing conditions as well as experimental validation.

## 6. Reflective questions

- How should clinicians balance dead-space elimination with the risk of overpacking deep cavity wounds?
- As demonstrated in this study, given the effects of wound filling ratio and the use of materials stiffer than adipose tissue on the viability of surrounding tissues, should absorptive capacity be reconsidered as a secondary property relative to mechanical behaviour ?
- How might future wound filler development reconcile the competing demands of mechanical tissue protection and dynamic wound environment variability?